\documentclass[english,review]{elsarticle}

\usepackage{lineno,hyperref}
\modulolinenumbers[5]

\journal{Physica Scripta}
\usepackage{xcolor}
\usepackage{ae,aecompl}
\usepackage[T1]{fontenc}
\usepackage[latin9]{inputenc}
\usepackage{babel}
\usepackage{amsmath}
\usepackage{amssymb}
\usepackage{graphicx}
\usepackage{lmodern}
\usepackage{amssymb,stackengine,graphicx}

\begin{document}
\selectlanguage{english}
\begin{frontmatter}

\title{Probing Band-center Anomaly with the Kernel Polynomial Method}
\author{N. A. Khan\corref{mycorrespondingauthor}}

\address{Centro de F\' isica das Universidades do Minho e Porto~\\
Departamento de F\' isica e Astronomia, Faculdade de Ci\^encias, Universidade
do Porto, 4169-007 Porto, Portugal\\ Department of Physics HITEC University, Cantt Taxila, Kalabagh-Nathia Gali Road, Taxila 47080 Punjab, Pakistan}
\cortext[mycorrespondingauthor]{Corresponding author}
\ead{niaz\_phy@yahoo.com}


\author{Syed Tahir Amin}
\address{Departamento de F\'isica, Instituto Superior T\'ecnico, Universidade de
Lisboa, Av. Rovisco Pais, 1049-001 Lisboa, Portugal \\ Departamento de Matem\'atica, Instituto Superior T\'ecnico, Universidade
de Lisboa, Av. Rovisco Pais, 1049-001 Lisboa, Portugal\\ CeFEMA, Instituto Superior T\'ecnico, Universidade de Lisboa, Av. Rovisco Pais, 1049-001 Lisboa, Portugal}



\begin{abstract}
We investigate the anomalous behavior of localization length of a non-interacting one-dimensional Anderson model at zero temperature. We report numerical calculations of the Thouless expression of localization length, based on the Kernel polynomial method (KPM), which has an $\mathcal{O}(N)$ computational complexity, where $N$ is the system size. The KPM results show excellent agreement with perturbative results in a large system size limit, confirming the validity of the Thouless formula. In the perturbative regime, we show that the KPM approximation of the Thouless expression produces the correct localization length at the band center in the thermodynamic limit.

The Thouless expression relates localization length in terms of density of states in a one-dimensional disordered system. By calculating the
KPM estimates of the density of states, we find a cusp-like behavior around the band center in the perturbative regime. This cusp-like singularity can not be obtained by approximate analytical calculations within the second-order approximations, reflects the band-center anomaly.
\end{abstract}

\end{frontmatter}



\section{Introduction}

It is a well-established fact that all eigenstates of a one-dimensional
(1D) non-interacting system is localized in the presence of infinitesimal
Anderson disorder at zero temperature $(T=0)$ \citep{MacKinnon1980,LeeRamakrishnan,Kramer1993}.
However, Anderson transitions may occur in a non-interacting cubic
lattice for a sufficiently large Anderson disorder (critical disorder)
at zero temperature  \citep{MacKinnon1980,MacKinnon1981}.
Typically electron states in the tails of the disorder broadened bands
can easily be restricted to a finite region of space, where the localization
length is much smaller than the system size. In addition, Mott and
Twose \citep{Mott1961}, proposed the idea of mobility edge which
separate localized and extended states.

The amplitudes of the one-particle wave-functions $\Psi(x)$, in 1D
Anderson model are exponentially decaying in space at infinity i.e.,
\begin{equation}
\left|\Psi(x)\right|=\exp(-\left|\vec{x}-\vec{x}_{0}\right|/\xi),
\end{equation}
in the limit $\left|\vec{x}-\vec{x}_{0}\right|\rightarrow\infty,$
where $\vec{x}_{0}$ is the center of localization. The parameter
$\xi$, known as the localization length, quantifies the localization
of wavefunctions~\citep{Thouless1972,Thouless1974,Thouless1979}.
It is one of the most fascinating tools used to characterize the behavior
of the disordered system. For exponentially localized states the
localization length is finite and typically smaller than the linear
size of the system. In contrast, the localization length tends to infinity
for extended states.

Traditionally, much interest has been given to the standard Anderson
model, both by rigorous analytical methods \citep{Thouless1972,Thouless1979,Izrailev1999}
and numerical simulations \citep{Kramer1993,Hatano2016}. Thouless
formula \citep{Thouless1972} relates the localization length $\xi$
of a 1D disordered electronic system in terms of the density of states
as given by
\begin{align}
\frac{1}{\xi(E)} & =\int\rho(\epsilon)\ln\left|E-\epsilon\right|\text{d}\epsilon-\ln\left|t\right|,\label{eq:ThoulessFormula}
\end{align}
where $\rho(\epsilon)$ is the density of states at energy $\epsilon$,
and $t$ is a hopping matrix element. Using perturbation theory to
second order, the localization length of the 1D Anderson model at
energy $E$ is written as \citep{Thouless1979,NiazPhDThesis}
\begin{equation}
\xi(E)=\frac{96t^{2}}{W^{2}}\left(1-\left(\frac{E}{2t}\right)^{2}\right),\label{eq:LLAndersonDisorder}
\end{equation}
in the weak disorder limit. The parameter $W$ is a positive constant
describing the strength of disorder. Expression \ref{eq:LLAndersonDisorder}
shows a power-law divergence of $\xi(E)$ in the limit of vanishing
disorder.

\emph{Band-center Anomaly:}  Czycholl et al. \citep{Czycholl1981} reported numerical calculations of the localization length based on the matrix inversion method, direct calculations, and mean a square extension of the Hamiltonian's eigenstates for the Anderson model in the perturbative regime. The numerical data were found in close agreement with Thouless overall energies \emph{except} at the vicinity of the band center $(E=0)$, referred to as ``band-center anomaly''. It is worthwhile to mention that the Thouless analytical estimate of localization length is $96/W^{2}$, whereas the numerical calculations derived with different approaches turn out to be $105.2/W^{2}$ at the band center in the weak disorder limit.  Kappus and Wegner~\citep{Kappus1981} showed that this band-center anomaly is a resonance effect that reflects the failure of the Born approximation. In addition, Derrida and Gardner~\citep{Derrida1984} verified the band-center anomaly and suggest that other anomalies should appear for resonant energies $E=2\cos(\alpha\pi)$, where $\alpha$ is a rational number.
Using the Hamiltonian map approach, Tessieri et.~al.,~\citep{TESSIERI2012}
derived an analytical expression of localization length for the 1D system
with diagonal disorder. This analytical treatment of localization
length was found to be in excellent agreement with numerical calculations
\citep{Czycholl1981}. The band center and band edges anomalies were
analytically investigated by Izrailev et.~al.,~\citep{Izrailev2012}.
The band center anomaly has also been investigated for the 1D model
with weak correlated disorder \citep{Tessieri_2015}. It was found
that the existence of disorder correlations can augment the band-center
anomaly. Recently, the anomalous behavior of localization length of
the 1D system with a specific kind of short-range correlated disorder
has been examined in \citep{Herrera2017}. The localization length
was found to increase with the third power of the correlation length.
More recently, the band-center anomaly of the system in the presence
of an imaginary random potential has been investigated \citep{Nguyen_2020}.
They have pointed out that the usual localization anomalies may be
strongly enhanced in the present non-Hermitian model.

The numerical calculations of localization length (measured from the scaling of the conductance, Lyapunov exponent, inverse participation ratio) for the 1D Anderson model revealed that the Thouless formula (Eq.~\ref{eq:ThoulessFormula})
fails to reproduce the correct localization length at the band center \citep{Czycholl1981,Kappus1981,Derrida1984,TESSIERI2012,Izrailev2012}.
In this paper, we investigate the band-center anomaly by numerically
calculating the localization length for the 1D Anderson model at the band center. In particular, we report numerical simulations of the Thouless formula of localization length, based on polynomial expansion technique the so-called Kernel Polynomial Method (KPM). The KPM estimates of localization length derived from the Thouless expression show an excellent agreement with the numerical data obtained from other numerical procedures, reflects that the KPM expansions of the Thouless formula can reproduce the correct localization length at the vicinity of the band center.
In addition, we calculate the KPM estimates of the density of states
for the model. The density of states at the band center show a cusp-like
singularity in the perturbative regime, that smooths out with increasing
disorder. We argue that the anomalous behavior of the localization
length at the band center or its neighborhood is due to this small
cusp-like behavior of the density of states.

The main focus is to examine the band center anomaly in the 1D Anderson model, which has been the subject of a long-standing debate. It is noted that the single parameter scaling (SPS) \citep{Schomerus2003} is violated for the model at the band-center as well as near the band-edges. The problem of SPS at the band center is resolved by Deych et. al., \citep{Deych2003}, who proposed that this spectral point splits the conduction band into two adjacent bands with a boundary, results in the violation of the single parameter scaling.

There exists a considerable amount of work dedicated to the numerical study of the strongly disordered electronic systems. The KPM turns out to be a powerful numerical method, which plays a very prominent
role in this research area \citep{Weibe2006,Weibe2008,Aires2015,Niaz2019}.
It is based on the polynomial expansions of the target function, which
yields high accuracy results with modest computational effort. The
Chebyshev polynomial \citep{Mason2002}---good convergence properties
of the corresponding series and close relation to Fourier transform---turns out to be a judicious choice of KPM expansion for most applications.
It is important to mention that the numerical convergence and accuracy
of the KPM estimates can be controlled by the number of polynomial
moments and optimal kernel.

The structure of the paper is as follows. In section~\ref{sec:Model-and-Computational},
we give a brief description of the 1D Anderson model and discuss the
kernel polynomial approximations of the Thouless expression and density
of states. In section~\ref{sec:Numerical-Results}, we discuss
our numerical results and explore the band-center anomaly. We discuss
the origin of the band-center anomaly and show that the kernel polynomial
approximations of the Thouless formula give very good estimates of
the data. In the last section, we sum up our conclusions.

\section{Model and Computational Method\label{sec:Model-and-Computational}}

\subsection{Anderson Model}

The lattice model under observation is a non-interacting tight-binding
system with nearest-neighbor hopping and random site energies. The
Hamiltonian in second quantization has the following general form,
\begin{figure}
\centering{}\includegraphics[scale=0.30]{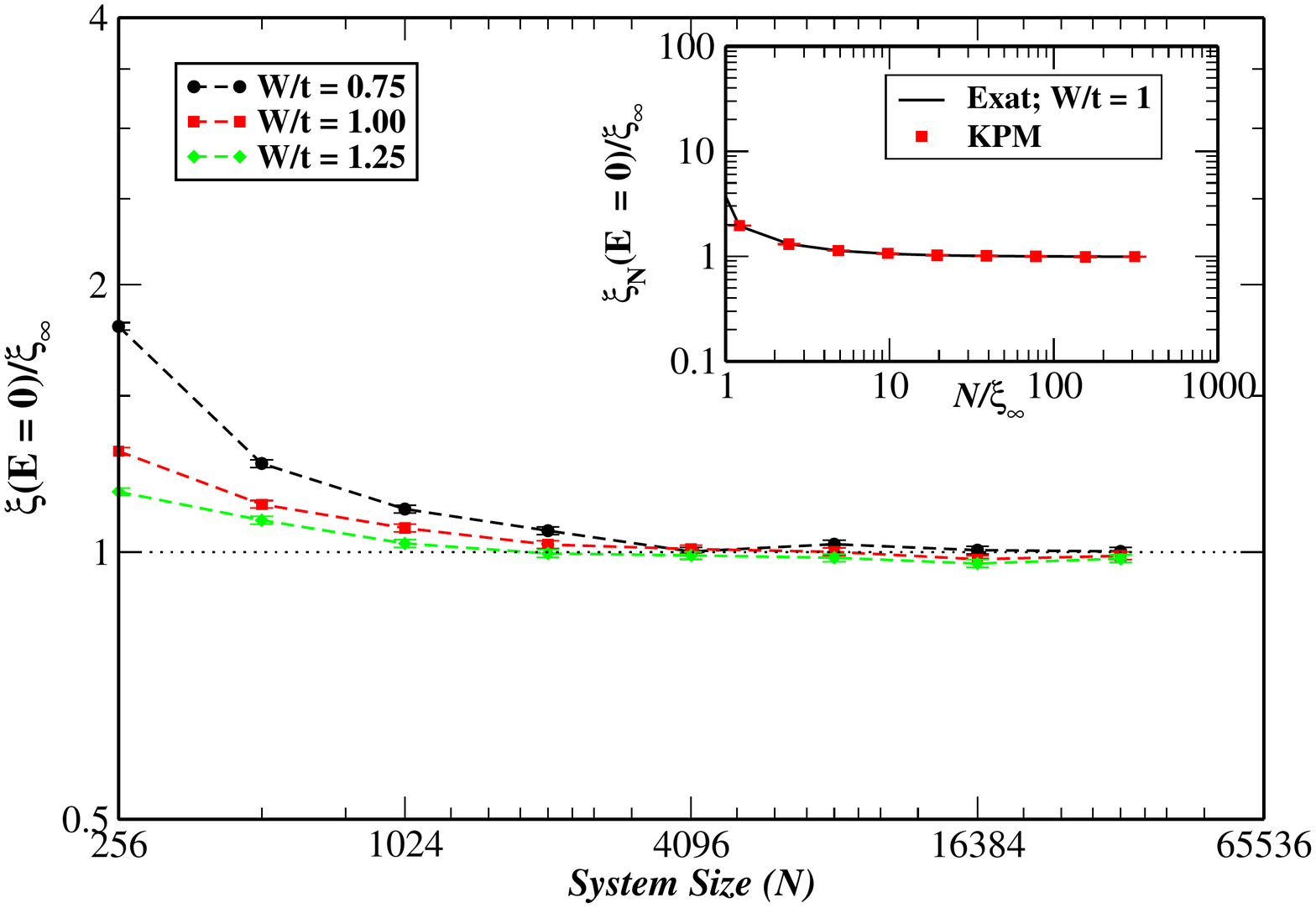}\caption{ Log-log plot of the KPM estimates of the rescaled localization length $\xi(E=0)/\xi_{\infty}$ as a function of system size $N$ for various disorder strength of the 1D Anderson model with OBC at zero temperature. The numerical calculations are carried out by using Eq.\ref{eq:LLKPM} with a fixed ($1\%$) estimated error (fluctuations in localization length) and $M=2048$ Chebyshev moments. The parameter, $\xi_{\infty} = 105.2/W^2$, is the analytically calculated localization length in the perturbative regime for $N\rightarrow\infty$ at $E = 0$ \citep{Izrailev2012}. The inset shows the log-log plot of the rescaled localization length $\xi(E=0)/\xi_{\infty}$ calculated by the KPM estimates and the exact diagonalization method of Thouless expression (Eq.\ref{eq:ThoulessFormula}) as a function of $N/\xi_{\infty}$ for $W = 1.0t$.}\label{fig:wLLcong}
\end{figure}
\begin{equation}
\mathcal{H}=-\sum_{\left\langle ij\right\rangle }t_{ij}(c_{i}^{\dagger}c_{j}+h.c.)+\sum_{i}\varepsilon_{i}c_{i}^{\dagger}c_{i}.\label{eq:1DHamiltonian}
\end{equation}
The essential parameters in the model are the transfer integrals $t$
and the diagonal disorder potentials $\varepsilon_{i}$. The transfer
integrals $t_{ij}=t=1,$ are restricted to nearest neighbors. All
energy scales are measured in a unit of $t$. For the Anderson model,
the on-site energies $\varepsilon_{i}$ are the independent random
variables uniformly distributed in the interval $[-\frac{W}{2},\,\frac{W}{2}]$,
where $W$ is the width of distribution which controls the strength
of disorder.

\subsection{Kernel Polynomial Method}

The kernel polynomial method (KPM) \citep{Weibe2006} is a polynomial
expansion-based technique and can efficiently compute the spectral
function of a large disordered quantum system. It has been successfully
applied to condensed matter problems \citep{Weibe2008,Aires2015,Niaz2019,Niaz2020},
for underpinning the Anderson transitions in non-interacting disordered
systems. In the KPM technique, the Hamiltonian and all energy scales
need to be normalized into the standard domain of orthogonality of
the Chebyshev polynomials $(\left]-1\,1\right[)$. Moreover, the numerical
convergence and accuracy of the KPM estimate strongly depend on the
Gibbs damping factor and coefficients of Chebyshev polynomials \citep{Weibe2006}.
The first kind of $m^{th}$ degree Chebyshev polynomials $T_{m}(x)$
are defined as
\begin{equation}
T_{m}(x)=\cos(m\,\arccos(x)),\qquad m\in\mathbb{N}.\label{eq:polynomial}
\end{equation}
 The $T_{m}(x)$ obey the following three-term recurrence relation
\begin{equation}
T_{m}(x)=2xT_{m-1}(x)-T_{m-2}(x),\qquad m>1,\label{eq:recurrsion}
\end{equation}
starting with $T_{0}(x)=1$ and $T_{1}(x)=x$; moreover it satisfy
the orthogonality relation 
\begin{equation}
\int_{-1}^{1}\,T_{n}(x)T_{m}(x)(1-x^{2})^{-1/2}dx=\frac{\pi}{2}\delta_{n,m}(\delta_{n,0}+1).\label{eq:Orthognal}
\end{equation}
The KPM estimate of the density of states \citep{Weibe2006} is, 
\begin{equation}
\rho(E)=\frac{2}{\pi\sqrt{1-E^{2}}}\sum_{m=0}^{M-1}\frac{g_{m}\mu_{m}}{(1+\delta_{m,0})}T_{m}(E),\label{eq:DOSkpm}
\end{equation}
where the expansion coefficients $\mu_{m}$ are determined as,
\begin{figure}
\centering{}\includegraphics[scale=0.30]{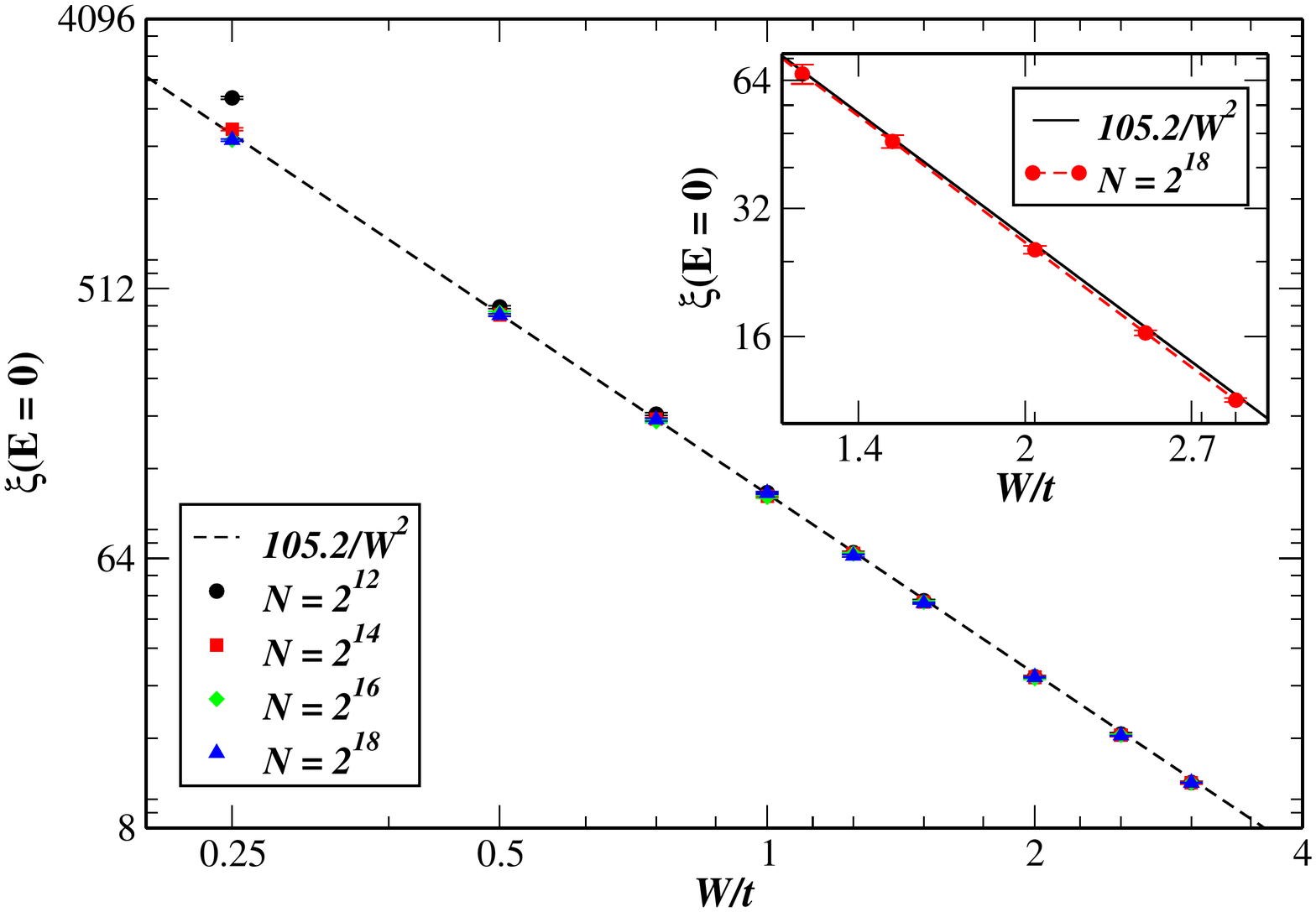}\caption{Log-log plot of the KPM estimates of localization length $\xi(E=0)$, as a function of $W/t$ for 1D Anderson model with open boundary conditions and $2048$ Chebyshev moments. A small deviation of the $\xi(E=0)$ from analytical result (black dashed line) appears in the small disorder limit, starts to converge with increasing system size. The inset shows the disorder scale that demarcates the weak and strong disorder regime. \label{fig:wLLKPM}}
\end{figure}
\begin{align}
\mu_{m} & =\int_{-1}^{1}T_{m}(E)\rho(E)\,dE,\nonumber \\
 & =\frac{1}{N}\text{Tr}[T_{m}(\hat{\mathcal{H}})].\label{eq:ChebM-1}
\end{align}
The trace in Eq.~\ref{eq:ChebM-1} can also be evaluated by the
stochastic evaluation method of traces (see Ref.~\citep{Weibe2006}
for the detail).

The expression in Eq.~\ref{eq:DOSkpm} represents the truncated sum
over $m$ Chebyshev series. This abrupt truncation of the series can
introduce unwanted oscillations, namely Gibbs oscillations, that can
be filtered out by employing an optimized damping factor. The most
favorable filter is the so-called Jackson Kernel \citep{Weibe2006}
$g_{m}$, defined as follows 
\begin{equation}
g_{m}=\frac{1}{M+1}((M-m+1)\cos (mx)+\sin (mx)\cot x),\label{eq:Jackson}
\end{equation}
with $x=\pi/(M+1)$. Here, $M$ is total number of Chebyshev moments.
The localization length of the model can be numerically calculated
by employing the kernel polynomial approximation of Thouless expression
Eq.~\ref{eq:ThoulessFormula}. As a result, the KPM estimates of
the localization length reads: 
\begin{equation}
\frac{1}{\xi(E)}=\frac{2}{\pi}\sum_{m=0}^{M-1}\frac{\mu_{m}g_{m}}{1+\delta_{m,0}}\mathcal{F}_{m}(E),\label{eq:ILL2}
\end{equation}
where the function $\mathcal{F}_{m}(E)$ is given by 
\begin{equation}
\mathcal{F}_{m}(E)=\int_{-1}^{1}T_{m}\left(\epsilon\right)\ln\left|E-\epsilon\right|\frac{\text{d\ensuremath{\epsilon}}}{\sqrt{1-\epsilon^{2}}}.\label{eq:int0}
\end{equation}
 The final expression of the localization length is written as \citep{Hatano2016}
\begin{equation}
\frac{1}{\xi(E)}=-\ln2-2\sum_{m=1}^{M-1}\frac{\mu_{m}g_{m}}{m}T_{m}(E).\label{eq:LLKPM}
\end{equation}
The above expression is used to compute the localization length of
the disordered systems for a given energy $E$. Here, we focus on
the KPM implementations of $\xi(E)$ at the band center.

\section{Numerical Results\label{sec:Numerical-Results}}
We numerically compute the KPM estimates of localization lenght (Eq.~\ref{eq:LLKPM}) as a function of system size for the non-interacting 1D Anderson model at absolute zero temperature as illustrated in Fig.~\ref{fig:wLLcong}. The KPM simulations are carried out for a fixed estimated error ($1\%$) with $M=2048$ Chebyshev moments. It is shown that the KPM estimates of localization length for stronger disorder strength converge faster than weaker disorder. Thus, we need to have a bigger system size to achieve better resolution for weaker disorder potential. In fact, one can clearly see the convergence of finite lattice results toward the perturbative thermodynamic limit result ($\xi_{\infty}$) with increasing size of the chain for various fixed disorder strength at the band center, $E=0$. In order to validate our KPM procedure, we compare the rescaled KPM estimates of localization length with the numerical data obtained by the exact diagonalization method as depicted in the Fig.~\ref{fig:wLLcong} (inset). Our finding shows that the KPM estimates are in good agreement with the results obtained by the exact diagonalization method, confirming the validity of the KPM procedure. It must be noted that the numerical complexity of the KPM procedure for the calculation of localization length is $\mathcal{O}(N)$ for a sparse Hamiltonian matrix, whereas the cost of solving a dense symmetric matrix require $\mathcal{O}(N^{3})$ operations. Thus, a straightforward diagonalization scheme for the eigenproblems is limited to small systems.

\begin{figure}
\centering{}\includegraphics[scale=0.34]{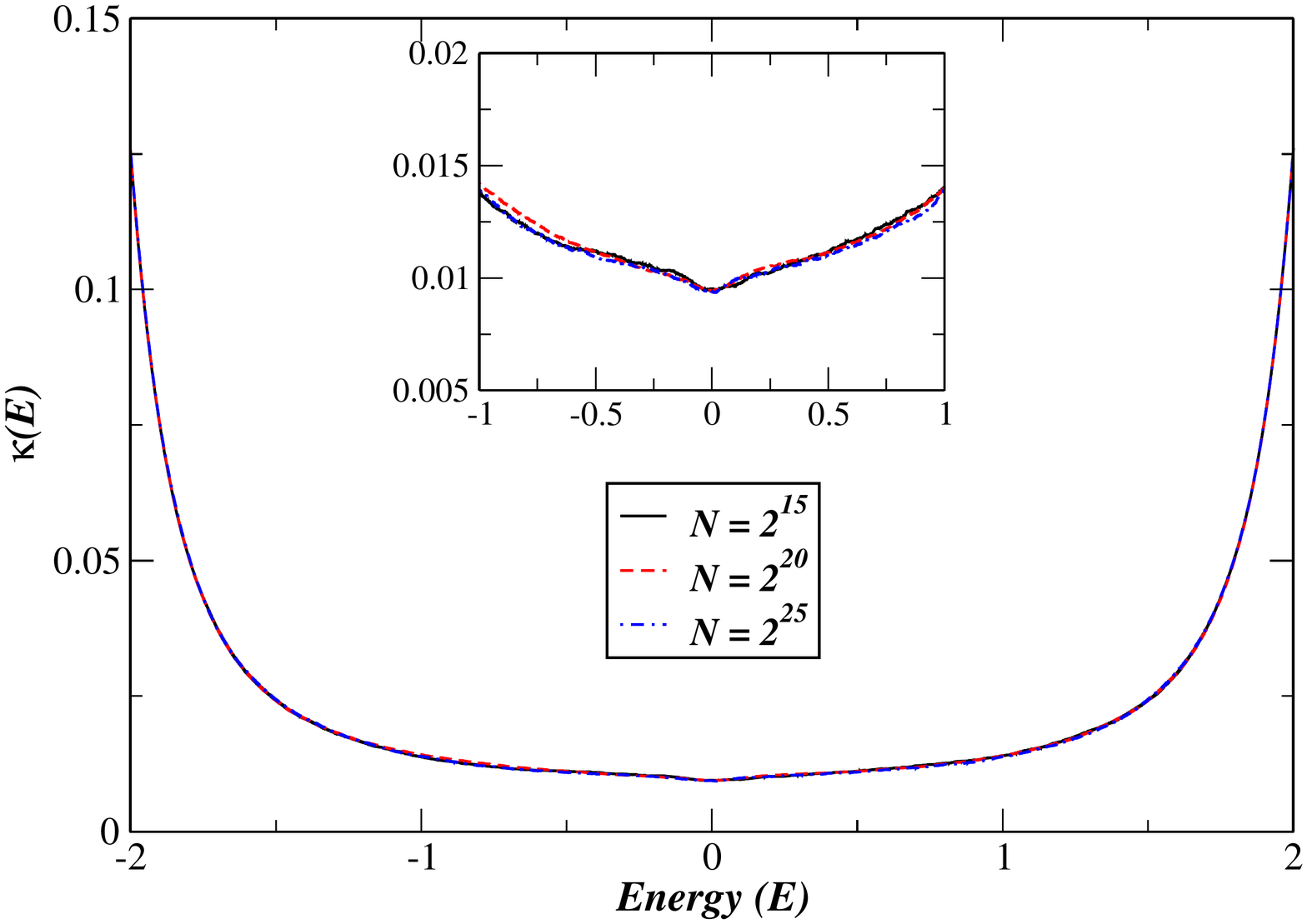}\caption{ The inverse localization length as a function of energy $E$ for various system size with $M = 16384$ Chebyshev polynomial moments,  $W = 1.0t$ disorder strength and averaged over $1024$ realizations of disorder. The inset shows the result near the band center.\label{fig:InverseLLvsE}}
\end{figure}
\begin{figure}
\centering{}\includegraphics[scale=0.34]{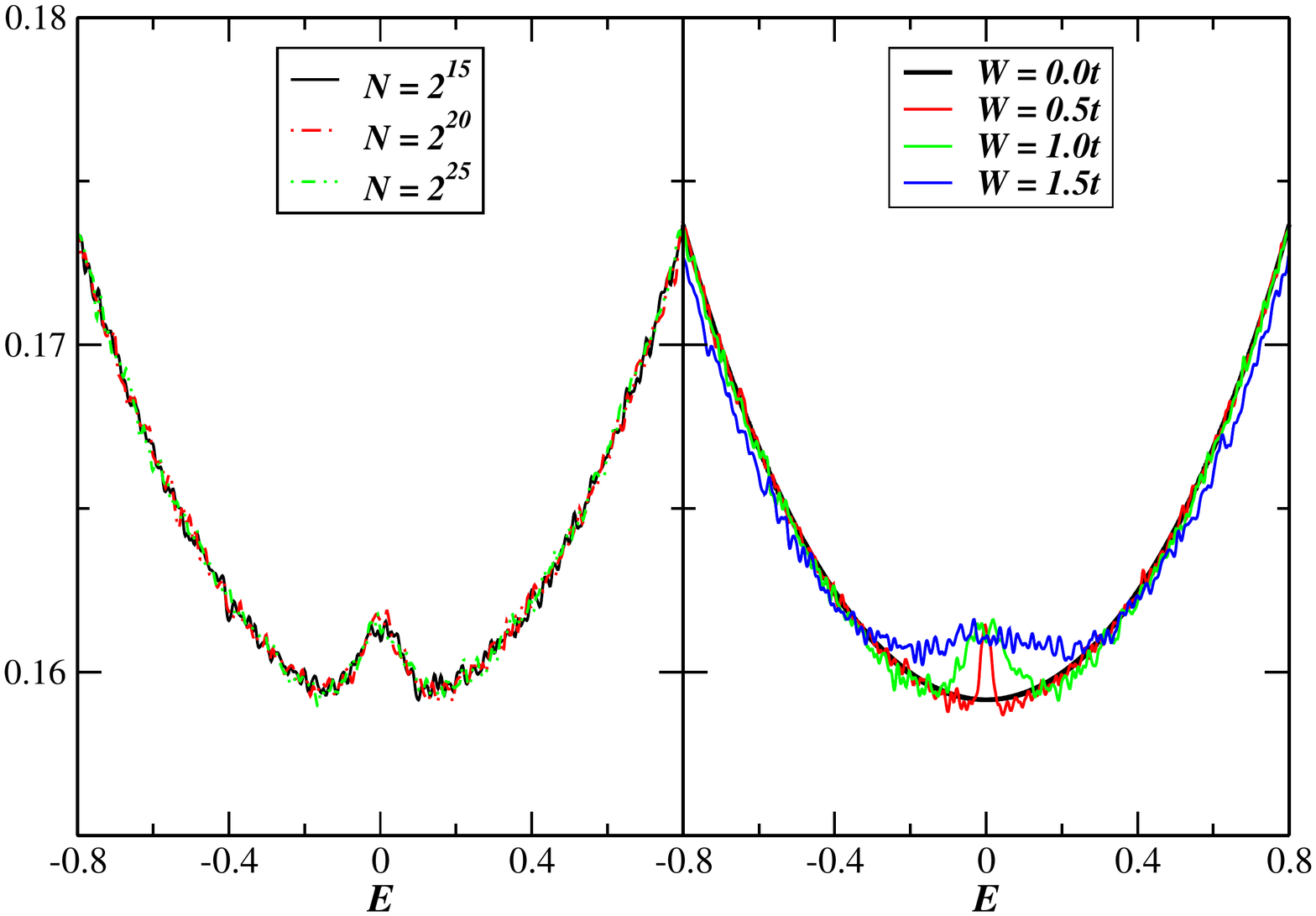}\caption{ The density of states as a function of energy $E$ of the 1D Anderson model for various system size with $M = 2048$ and $W = 1.0t$ (left panel) and disorder strength with $N = 2^{20}$ and $M = 2048$ (right panel). All the computations are carried out for the system of size $N = 2^n$ and averaged over $2^{31-n}$ realizations of disorder with $n \in\mathbb{N}$.\label{fig:1DDosAndersonModel}}
\end{figure}

Our main goal is to compute the localization length and to study the
described discrepancy about the anomalous behavior of localization
length at the band center. According to the literature, the Thouless
expression of localization length fails to reproduce the expected
result \citep{Czycholl1981,Kappus1981,Derrida1984,TESSIERI2012,Izrailev2012,Tessieri_2015,Herrera2017,Nguyen_2020}.
However, our numerical findings, based on the kernel polynomial expansions of Thouless expression revealed the correct localization length for
the 1D Anderson model at the band center as shown in Fig.~\ref{fig:wLLKPM}.
The numerical calculations are carried out for $M=2048$ Chebyshev
moments with a fixed estimated error of $1\%$ for a different size of
the system at the band center. The estimated error is determined by
the fluctuations in localization length. As shown in Fig.~\ref{fig:wLLKPM}, a small deviation of the estimated localization length is observed
for $N=2^{12}$ in the limit of small disorder strength, which starts
to disappear with increasing system size. It is because one needs
to have a larger system size for better resolutions as illustrated in Fig.~\ref{fig:wLLcong}. Hence, the KPM approximations of the Thouless formula yield the expected localization length for the Anderson model at the band center and is in good agreement with the perturbative analytical \citep{Izrailev2012} and numerical \citep{Czycholl1981} results in the limit of large system size. In Fig.~\ref{fig:wLLKPM} (inset), we predict a boundary between strong and weak disorder strength. It is shown that the perturbative result give a good account of the KPM estimates of  localization length for the disorder of strength $W\lesssim 2.0t$ with variance of the local disorder potential $\sigma_{\varepsilon}^{2}=W^{2}/12\lesssim 1/3$ in the thermodynamic limit. This result is consistent with the data presented in \citep{Niaz2019}, where the disorder-averaged spectral function follows Lorentzian distributions in the perturbative regime whereas for strong disorder the spectral function is not very well fitted to Lorentzian function.

Let us now turn to the situation where the inverse localization length, $\kappa(E)$, vary as a function of energy in small disorder limit. For instance, we calculate the KPM estimates of $\kappa(E)$ for various size of the system with disorder $W=1.0t$ and $M = 16384$ polynomial moments in the entire energy range as depicted in Fig.~\ref{fig:InverseLLvsE}. We observe a dip shape of $\kappa(E)$  around the band center for each size of system, is consistent with the results presented in \citep{TESSIERI2012}. The inset shows a clear view of the result (dip-like feature of the $\kappa(E)$) in the neighborhood of the band center. This dip of the $\kappa(E)$ is due to the cusp-like behavior of the density of states around the band center, which can not be captured by using second order perturbation theory \citep{Thouless1972, NiazPhDThesis}

To know the origin of the band-center anomaly of the Thouless expression of localization length, we examine the density of states of the Anderson model restricted to nearest-neighbor interactions as illustrated in Fig.~\ref{fig:1DDosAndersonModel}. It is important to mention that the density of states of the one-dimensional clean system (in the absence of disorder and interactions) is symmetric around the band center and exhibit van Hove singularities of the form of a square root divergence at the band edges. However, these band edges singularities progressively broadened with disorder strength and have been vanished in the strong disorder limit \citep{Thouless1974,Kappus1981}. Here, we are interested to study the density of states of the 1D Anderson model around the band center at zero temperature. Therefore, we compute the density of states of the model in the neighborhood of the band center. All the numerical calculations are based on the KPM technique, carried out for the system of size $N = 2^n$ and averaged over $2^{31-n}$ realizations of disorder with $n \in\mathbb{N}$. We first test and confirm the numerical convergence of the KPM estimates of density of states for various system sizes as presented in the left panel of Fig.~\ref{fig:1DDosAndersonModel}. We have observed a size-independent nature of the density of states in the large system size and small disorder limit. This result is consistent with the results presented in \citep{Rossum1994}, where the density of states is shown to be self-averaging in the perturbative regime. In the right panel of Fig.~\ref{fig:1DDosAndersonModel}, we demonstrated the roles of disorder effects on the density of states for fixed system size and moments. One can clearly see a sharp peak of the density of states at the band center in the presence of weak disorder. This sharp peak progressively broaden with increasing disorder and, consequently, becomes flat in the limit of strong disorder.
In fact, the peak has a constant fixed value for a variety of disorder strengths. It is basically the density of states of the neighborhood of $E=0$ that starts to increase with disorder strength, and eventually, the sharp peak disappears in the strong disorder limit. It is noted that the KPM estimate of the density of states is approximately $0.16123$ at the band center in the limit of large system of size ($N=2^{25}$). Moreover, using degenerate perturbation theory, the approximated density of states $\rho(E)$, at energy $E$ reduced to \citep{Kappus1981}
\begin{equation}
\rho(E)=\frac{1}{2\pi t}\left(1+\frac{W^{4}\left(W^{2}-(8Et)^{2}\right)}{72\left(W^{4}+(8Et)^{2}\right)^{2}}\right).\label{eq:DPT}
\end{equation}
At the band center, $E=0$, the approximated density of states turns out to be $0.16136$. Moreover, Derrida and Gardner \citep{Derrida1984} have proved that the density of states at $E=0$ is $0.16156$. This ensure that the KPM technique yields a very good estimates of the data (cusp-like feature of the density of states), in an excellent agreement with the result presented in \citep{Kappus1981,Derrida1984}. Furthermore, it is figured out that the cusp-like behavior of density of states in the vicinity of the band center is independent of the disorder strength. As we know that the Thouless expression (Eq.~\ref{eq:ThoulessFormula}) connects inverse localization length and density of states of a one-dimensional disordered model \citep{Thouless1972}. Therefore, one may expect similar anomalous behavior of the inverse localization length. Indeed, we obtain a dip of the inverse localization length in the vicinity of band center. It is worth mentioning that, this dip like behavior of inverse localization length is due to the small cusp singularity of the density of state around the band center, non-analytical within the second-order perturbation theory, reflects the anomalous behavior of the localization length.

\section{Conclusions\label{sec:Conclusions}}

We have investigated the anomalous behavior of localization length
for the non-interacting 1D Anderson model at zero temperature. Contrary
to the previous numerical and analytical results \citep{Czycholl1981,Kappus1981},
we have figured out that the kernel polynomial approximations of
Thouless expression can reproduce the correct or expected localization
length at the band center. However, the analytical treatment of the
Thouless formula fails at the band center (due to the failure of the
Born approximation).

The Thouless formula connects localization length and density of states. Hence, similar anomalous behavior has been observed for the density of states. In fact, we have noticed a cusp-like behavior of the density of states in the vicinity of the band center in the perturbative regime.
This sharp-peaked density of states can not be encountered using perturbation theory to second order. We argued that this cusp-like singularity of the density of states around the band center is the reason for the anomalous behavior of the localization length.

\section*{Acknowledgments}
I wish to express my sincere gratitude to Prof. J. M. B. Lopes dos Santos for invaluable guidance, helpful suggestions and Prof. J. M. Viana Parente Lopes for teaching me a most fascinating numerical method, ``Kernel Polynomial Method''.

For this work, NAK was supported by the INTERWEAVE project, Erasmus
Mundus Action 2 Strand 1 Lot 11, EACEA/42/11 Grant Agreement 2013-2538/001-001
EM Action 2 Partnership Asia-Europe, Funda\c{c}\~ao da Ci\^encia e Tecnologia
and COMPETE 2020 program in FEDER component (EU), through the projects
POCI-01-0145-FEDER-028887 and UID/FIS/04650/2013.STA thank the support from
the Funda\c{c}\~ao da Ci\^encia e Tecnologia (FCT) through Doctoral Programme in the Physics and
Mathematics of Information and the associated scholarship PD/ BD/113651/2015 and through the grant UID/CTM/04540/2013. STA also gratefully
acknowledges the support of SQIG -{}- Security and Quantum Information
Group, under the Funda\c{c}\~ao para a Ci\^encia e a Tecnologia (FCT) project
UID/EEA/50008/2019, and European funds, namely H2020 project SPARTA.


\bibliography{BCAKPM}

\end{document}